\documentstyle[pre,preprint,aps,tighten]{revtex}

\begin{document}
\title{Dynamical phase diagram of the dc-driven underdamped Frenkel-Kontorova chain}
\author{Maxim~Paliy~$^{\dag, \ddagger}$, Oleg~Braun~$^{\ddagger}$,
Thierry~Dauxois$^{\S}$, and Bambi~Hu$^{\dag, \diamond}$}
\address{$^{\dag}$
Centre for Nonlinear Studies and Department of Physics, \\
Hong Kong Baptist University, Kowloon Tong, Hong Kong}
\address{$^{\ddagger}$ Institute of Physics,\\
Ukrainian Academy of Sciences,\\
46 Science Avenue, UA-252022 Kiev, Ukraine}
\address{$^{\S}$ Laboratoire de Physique,\\
Ecole Normale Sup\'{e}rieure de Lyon,\\
46 All\'{e}e d'Italie, 69364 Lyon C\'{e}dex 07, France}
\address{$^{\diamond}$ Department of Physics,\\
University of Houston, Houston, TX 77204, USA}
\date{\today}
\maketitle

\begin{abstract}
Multistep dynamical phase transition from the locked to the running state of
atoms in response to a dc external force is studied by MD simulations of the
generalized Frenkel-Kontorova model in the underdamped limit. We show that
the hierarchy of transition recently reported [Braun et al, Phys. Rev. Lett.
{\bf 78}, 1295 (1997)]  strongly depends on the value of the friction
constant. A simple phenomenological explanation for the friction dependence
of the various critical forces separating intermediate regimes is given.
\end{abstract}

\pacs{PACSnumbers: 05.70.Ln; 46.10.+z; 66.30.-h; 63.20.Ry}

\section{Introduction}

\label{s1}

The nonlinear response of a system of interacting atoms to a dc driving
force has recently attracted a great interest (see \cite
{Trullinger1978,GuyerMiller1978b,ButtikerLandauer1981,Marchesoni,ariyasu87,persson,avalan,Family,LFloria,nl-mob}
and references therein). The knowledge of the {\em microscopic} mechanisms
for mobility, friction and lubrication processes are very important, in
particular, for a better understanding of solid friction at {\em macroscopic
level}, as well as in various fields of applied science and technology such
as adhesion, contact formation,
friction wear, lubrication, fracture, {\em etc}.

One generic example represents a layer of atoms adsorbed on a crystalline
surface. The adsorbate in this case is considered as an atomic sub-system
and the remainder is modeled as an external potential, a damping constant
and a thermal bath. Such a system can be treated within the framework of a
generalized Frenkel-Kontorova (FK) model \cite{fk,lnp,bklast}. When an
external dc force is applied to such a system, its response can be very
nonlinear and complex. By contrast, the driven motion of a single Brownian
particle in the external periodic potential has been studied in detail and
is now well understood \cite{Risken1984}. If the force $F$ is applied to the
particle, the total external potential in the direction $x$ of the force is
a corrugated plane, with a slope $F$. For small forces the potential has
local minima, and the particle is {\em locked}. The local minima disappear
at forces higher then $F_{f0}\equiv C \pi \varepsilon/a$, where $\varepsilon$
is the amplitude of the periodic potential, $a$ its period, and $C$ a
numerical factor depending on the shape of the potential. Thus, when the
applied force is adiabatically increased, the atom passes from the locked to
the {\em running} state at $F_{f0}$, and the mobility $B=\langle v \rangle /F
$ (where $\langle v \rangle$ is the drift velocity) reaches its maximal
value $B_f\equiv 1/m\eta$, where $m$ is the atomic mass and $\eta$ the
friction coefficient. On the other hand, if one decreases the force $F$
adiabatically starting from the running state, the critical force $F_{b0}
\simeq 4\eta \sqrt{m \varepsilon}/\pi$ for the backward transition to the
locked state is different owing to inertia of the system. Namely, in the
underdamped limit ($\eta \ll \omega_0$, where $\omega_0$ is the frequency of
atomic vibration in the external potential), the inequality $F_{b0} < F_{f0}$
holds, and one can observe {\em hysteresis}: in the {\em bistable} region
$F_{b0}<F<F_{f0}$, the particle is either locked or running depending on its
initial velocity. For a single particle, the bistability disappears at any
nonzero temperature. Besides, since the ``forward'' critical force $F_{f0}$
is independent on the friction, and the ``backward'' force $F_{b0}$ grows
linearly with friction, the width of hysteresis vanishes at $\eta > \eta^*
\equiv \frac{C \pi^2}{4 a} \sqrt{\frac{\varepsilon}{m}}$.

The problem of {\em interacting} particles in a periodic potential is much
more difficult. For the overdamped case ($\eta \gg \omega_0$) the nonlinear
mobility of the FK model has been studied in a number of papers \cite
{Trullinger1978,GuyerMiller1978b,ButtikerLandauer1981,Marchesoni,LFloria}.
By contrast, investigations of the underdamped case are very limited. In
that context, Persson~\cite{persson} observed {\em hysteretic} dynamical
phase transition, similar to the $T=0$ one-particle case, in the MD
simulation of a 2D system of interacting atoms subjected to a periodic
potential. Besides, our recent work \cite{nl-mob} on the underdamped
generalized FK model revealed strong collective effects in the dynamics of
the dc-force driven layer of atoms. When the external force increases, the
FK system exhibits a complex hierarchy of first-order dynamical phase
transitions from the completely immobile state to the totally running state,
passing through several ``slip-stick'' intermediate stages characterized by
the running state of collective {\em quasiparticle} excitations of the FK
model known as kinks \cite{nl-mob}. The scenario of these intermediate
transitions depends on whether the concentration of atoms corresponds to
trivial or complex atomic structure. All the observed transitions are
hysteretic and, it is remarkable, that, by contrast with the case of
noninteracting atoms, the hysteresis survives at {\em nonzero} temperature
of the system.

However, the results \cite{nl-mob} have been obtained for one value of the
friction constant $\eta $ only. That is why, in the present work we are
interested in the question how the observed dynamical transitions evolve
when the friction changed. We calculate dynamical ``phase diagrams'' of the
FK system in the $(F,\eta )$ plane for two generic atomic concentrations
(Sect.~\ref{rectangular}). We show that several critical forces, separating
intermediate stages during the transition from the locked state to the
running state of the atoms are friction dependent, and we propose a simple
phenomenological approach (Sect. \ref{pheno}) which allows to explain these
dependences.

\section{The model}

\label{model}

The detailed description of the generalized FK model under study, of the
numerical procedure, and also the substantiation of the choice of model
parameters can be found in Ref. \cite{nl-mob}; here we only outline the main
aspects of the model. The atomic motion is governed by the Langevin
equation,
\begin{equation}
\label{langevin}m\ddot{x_i} + m \eta \dot{x_i} + \frac{d}{d x_i} \left[
V_{sub}(x_i,y_i,z_i) + \sum_{j(j\neq i)} V_{int}(|\vec{r}_i-\vec{r}_j|)
\right]=F^{(x)} + \delta\!F_{i}^{(x)} (t),
\end{equation}
for the $x$ coordinate of $i$-th atom, and similar equations for $y$ and 
$z$. Here $m$ is the atomic mass, $V_{sub}(x,y,z)$ the external substrate
potential, $V_{int}(r)$ the potential of pairwise interaction between atoms,
$\eta$ the viscous friction constant, which corresponds to the rate of
energy exchange with the substrate, $\vec{F}=\{ F,0,0 \}$ the dc driving
force, and $\vec{\delta\!F}$ the Gaussian random force with correlation
function
\begin{equation}
\label{random}\langle \delta\!F^{(\alpha)}_i(t) \;  \delta\!F^{(\beta)}_j
(t^{\prime}) \rangle = 2 \eta m k_B T \delta_{\alpha \beta} \delta_{ij}
\delta(t-t^{\prime}).
\end{equation}

For a better representation of the natural microscopic scale of the problem,
we use a dimensional system of units, measuring distance in Angstr\"{o}ms,
energy and temperature in Electronvolts. The mass of atoms is chosen as
unity: $m=1$. In the remainder of the paper, the units of other dimensional
physical quantities are omitted, but they are expressed in terms of the
above units.

We use exponential interactions between atoms corresponding to the repulsion
between atomic cores,
\begin{equation}
\label{int}V_{int}(r) = V_0 \exp (-\beta_0 r),
\end{equation}
where $V_0 = 10$ eV is the amplitude and inverse of 
$\beta_0 = 0.85$ \AA $^{-1}$ determines the range of interaction.

To model the substrate, we used in the simulation the true 3D external
potential, periodic in the $(x,y)$ plane (with the rectangular symmetry) and
parabolic in the $z$ direction,
\begin{equation}
\label{sub}V_{sub}(x,y,z)=V_{pr}(x;a_{sx},\varepsilon_{sx},s_{x})
+V_{pr}(y;a_{sy},\varepsilon_{sy},s_{y})  +\frac{1}{2} m \omega_z^2 z^2,
\end{equation}
where $\omega_z$ is the frequency of normal vibration of a single atom, and
\begin{equation}
\label{pr}V_{pr}(x;a,\varepsilon,s) = \frac{1}{2} \varepsilon 
\frac {(1+s)^2[1-\cos (2\pi x/a)]} {1+s^2-2s\cos (2\pi x/a)}
\end{equation}
is the deformable Peyrard-Remoissenet potential \cite{peyrard-remoissenet},
and the parameter $s$, $|s| < 1$, describes its shape. The choice of the
lattice constants, $a_{sx}= 2.74$~\AA\, and $a_{sy}= 4.47$~\AA\ , and of the
energy barriers, $\varepsilon_{sx} = 0.46$~eV and 
$\varepsilon_{sy}=0.76$~eV, 
provide a high anisotropy of this potential, which can be viewed as the
set of ``channels'' with the corrugated bottoms, oriented along the $x$
direction. This potential is typical for the furrowed crystal surfaces,
namely our parameters were chosen for the Na-W(112) adsystem \cite{nl-mob}.
The frequency $\omega_x$ of a single-atom vibration along the $x$ direction
is connected to the shape parameter $s_x$ by the relationship $\omega_x =
\omega_0 (1+s_x)/(1-s_x)$, where $\omega_0 \equiv (\varepsilon_{sx}/2
m)^{1/2} (2 \pi/a_{sx})$. We have chosen typical values $s_x=0.2$ and
$s_y=0.4$, which lead to the frequencies of atomic vibrations $\omega_x=1.65$
and $\omega_y=2.02$ respectively. Finally, we took $\omega_z=\frac{1}{2}
(\omega_x + \omega_y) =1.84$. Note, that our choice of parameters does not
claim to be a quantitative interpretation of the concrete adsystem, because
the model is oversimplified. However, we do believe on a qualitative
description of the effect under investigation and claim that typical
adsystems should exhibit similar behaviors.

In the present work we study the behavior of the system in a wide range of
frictions in the underdamped limit $\eta \ll \omega_x$, corresponding to
typical adsystems \cite{fric}. In the simulation, we first look for the
minimum-energy configuration of the system. Then, we adiabatically increase
temperature and force and measure the mobility $B$ for given values $T$ and 
$F$ (this procedure was described in detail in \cite{nl-mob}). In order to
emphasize the phase transitions, the system is studied at a very low
substrate temperature, $T=0.0005$~eV.

An important parameter of the generalized FK model is the atomic
concentration. For the repulsive interatomic interaction used in the present
work, we have to impose the periodic boundary conditions in the $x$ and $y$
directions in order to fix the concentration. Namely, we place $N$ atoms
into the fixed area $L_x\times L_y$, where $L_x=M_xa_{sx}$ and $L_y=M_ya_{sy}
$, so that the dimensionless atomic concentration (the so-called coverage in
surface physics) is equal to $\theta =N/M$ ($M=M_xM_y$). The atomic
concentration in the FK system plays a crucial role since it defines the
number of quasiparticle excitations, i.e. the number of geometrical
(residual) kinks. These excitations can be defined for any background 
{\em commensurate} atomic structure $\theta _0=p/q$, where $p$ and $q$ are
relative prime integers \cite{bklast,braunKZ}. If the concentration $\theta $
slightly deviates from the background value $\theta _0$, the ground state of
the system corresponds to large domains with background commensurate
coverage $\theta _0$, separated by localized incommensurate zones of
compression (expansion) called kinks (antikinks). When the background
commensurate coverage is {\em trivial}, $\theta _0=1/q$, the kinks defined
on this structure are called {\em trivial} kinks \cite{bklast}. Besides, a
nontrivial background coverage $\theta _0=p/q$ ($p\neq 1$, with complex
elementary cells consisting of $p$ atoms) can be represented as a lattice of
trivial kinks, defined on the background of the closest trivial structure.
Therefore, in the latter case a deviation from the $\theta _0=p/q$ structure
can be represented as a discommensuration in the lattice of trivial kinks,
i.e.\ a ``kink in the kink lattice'', called {\em superkink} \cite{bklast}.

As in the simulation we study finite systems, we have to chose an
appropriate system size to insert $N_k$ kinks into the $\theta_0=p/q$
commensurate background structure; the integers $N$ and $M$ must satisfy the
equation \cite{braunKZ}
\begin{equation}
\label{NM}q N = p M + N_k \sigma \, ,
\end{equation}
where the topological charge $\sigma$ is $\sigma=+1$ for the kink and 
$\sigma=-1$ for the antikink. As background structures, we discuss here two
interesting cases, the trivial coverage $\theta_0=1/2$, and the complex
coverage $\theta_0=2/3$.

\section{Simulation results.}

\label{rectangular}

To simplify the problem, in the present work we only consider the
quasi-one-dimensional case, putting $M_y=1$, so that all chains move in the
same way (however, the interaction between the atoms, as well as the atomic
motion still has the 3D character). Let us note here, that this simplified
choice leads only to a minor difference in system behavior in comparison
with true 2D FK system with $M_y >1$ \cite{nl-mob}.  Namely, the exact
critical force depends slightly on the external conditions, which means that
the transitions do not occur simultaneously in all the chains. A careful
examination of the behavior in different chains shows an enhanced transition
to sliding state due to cooperative effects in the second dimension with an
exponential law \cite{nl-mob}. On the contrary, the transition of a chain to
the locked state is almost independent of neighboring chains.

\subsection{Trivial $\theta_0=1/2$ background coverage}

\label{rec-onehalf}

First, let us consider the ground state which corresponds to domains of the
trivial $\theta_0=1/2$ coverage separated by trivial kinks: $\theta=21/40$.
Namely, we took $N=105$ and $M_x=200$, having thus ten kinks over the length
under investigation with an average spacing of $20 a_{sx}$ between kinks.

The generic $B(F)$ dependence for the friction constant $\eta = 0.12 \omega_x
$ is presented in Fig.~\ref{rbf}(a). In this figure, as well as in all
figures below, it will be convenient to express force $F$ in the units of
the constant ``forward'' critical force for a single particle $F_{f0} = C
\pi \varepsilon_{sx}/a_{sx} \approx 0.607$ ($C \approx 1.15$ for the shape
of the external potential used in the present simulation). During the
force-increasing process, one can distinguish several steps in the $B(F)$
dependence, which correspond to a hierarchy of depinnings of kinks \cite
{nl-mob}. Namely, at $F<F_{tk} \approx 0.23F_{f0}$ the system is in the
completely immobile state, while at $F>F_{tk}$ the system has nonzero
mobility $B_{tk}$ due to the {\em running state} of trivial kinks (Fig.~\ref
{jam}~a). It was shown in \cite{nl-mob}, that the force $F_{tk}$ is related
to the degradation of the periodic Peierls-Nabarro potential 
$\varepsilon_{PN}$ for the trivial kinks: the kinks start to slide at 
$F>F_{tk} \approx \pi \varepsilon_{PN}/a_{ax}$. The second abrupt increase of
the mobility to the value $B_m$ takes place when the force exceeds a certain
threshold $F_{pair}\approx 0.35 F_{f0}$, connected with the vanishing of the
energy barrier for creation of additional kink-antikink pairs in the system,
so the number of mass carriers in the system increases, leading to the
increase of the mobility. The remarkable tendency for the running kinks is
to come closer to each other, thus bunching into compact groups. This
tendency is especially enhanced after the second threshold $F_{pair}$, where
the concentration of kinks is large. The bunched kinks build up dense groups
of immobile atoms with $\theta=1$, while the rest of the system consists of
running atoms (corresponding to the running state of force-excited
antikinks). This state, which is very reminiscent of a {\em traffic jam}
(Fig.~\ref{jam}(b)), survives till the last threshold force $F_f \approx
0.53F_{f0}$, after which all the atoms are sliding over the periodic
potential, and the system reaches the highest possible value of the mobility
$B_f=1/m\eta$.

During the force-decreasing process, the system jumps back to the immobile
state at the critical force $F_b$, and one can see a large hysteresis which,
contrary to the one-particle case, survives at nonzero temperature \cite
{nl-mob}.

As was mentioned in the Introduction, for a single Brownian particle the
``forward'' force $F_{f0}$ is independent on the friction, while the
``backward'' force $F_{b0}$ is linearly proportional to the friction. By
contrast,  the situation is more complex in the case of interacting atoms.
Fig.~\ref{rec12} represents the dynamical phase diagram in the $(F, \eta)$
plane, where we plot the critical forces $F_{tk}$, $F_{pair}$, $F_f$, and 
$F_b$ versus the friction coefficient $\eta$.

Let us consider first the forward transition from the locked to the running
state. One can distinguish two regions of friction corresponding to
different scenarios of the transition.

{\bf (i)} At very low frictions, $\eta < 0.05 \omega_x$, there is no
intermediate stages. When the force increases, the system jumps from the
locked to the running state directly at the force $F_f$, this force being
exactly equal to the critical force $F_{tk}$ for the {\em kink} transition
to the running state.

{\bf (ii)} By contrast, at larger frictions, $\eta > 0.05 \omega_x$, all the
above-mentioned intermediate stages with running kinks do exist. The second
difference from the very low friction region is that the ``forward'' force 
$F_f$ and the kink-antikink nucleation force $F_{pair}$ are 
{\em friction-dependent}, namely they grow with friction increasing, while the
critical ``kink'' force $F_{tk}$ remains constant. At not too high
frictions,  the critical force $F_f$ first increases approximately linearly
with the friction $\eta$, while at higher frictions the increase of $F_f$
slows down, and finally tends to a constant value $F_{f}^{\ast}$, close to
the value $F_{f0}$ for noninteracting atoms (Fig.~\ref{rec12}).

Unfortunately, we failed to plot the critical force $F_{pair}$ in Fig.~\ref
{rec12} at $\eta > 0.14 \omega _x$, because the ``traffic jam'' regime is
not well defined anymore at these higher frictions, probably because of the
small number of atoms in the immobile group (Fig. \ref{jam}(b)) at 
$\theta=21/40$ for the finite-size system. However, at a higher concentration
$\theta = 21/31$ (see next subsection \ref{rec-twothird}), where the
immobile groups are larger, the ``traffic jam'' regime is well defined at
all studied frictions.

The backward transition from the running to the locked state has one
interesting feature. Namely, the ``backward'' critical force $F_b$ grows
with $\eta$ increasing (notice, that at not too high frictions, $F_b(\eta)$
exactly matches the law $F_{b0} \simeq 4\eta \sqrt{m \varepsilon_s}/\pi$ for
{\em noninteracting} atoms, shown by the dashed line in Fig.~\ref{rec12}).
If the friction is high enough, $F_b$ is larger than the critical force for
kinks $F_{tk} = $const. Therefore, at $\eta < 0.1 \omega_x$ the system jumps
back to the completely {\em locked} state of atoms, while at $\eta > 0.1
\omega_x$ the backward transition has also multistep character: when the
force decreases, it first occurs a drop of the mobility $B$ from $B_f$ to
kink-mobility value $B_{tk}$, and only at some $F<F_{tk}$ the mobility
vanishes finally.

\subsection{Complex $\theta_0=2/3$ background coverage}

\label{rec-twothird}

Now let us describe the behavior of the system at the concentration 
$\theta=21/31$ (we used $N=105$ and $M_x=155$). The ground state in this case
corresponds to domains of complex $\theta_0=2/3$ commensurate structure,
separated by {\em superkinks} with an average spacing $30 a_{sx}$ between
them. On the other hand, the $\theta=2/3$ structure can be viewed as a dense
lattice of {\em trivial} kinks defined on the background of the $\theta_0=1/2
$ structure.

This specificity clearly manifests itself in the $B(F)$ dependence plotted
in Fig.~\ref{rbf}(b) for the same friction constant $\eta = 0.12 \omega_x$
as in Fig.~\ref{rbf}(a). During the force-increasing process, one can
distinguish two sharp steps of increasing of the mobility $B$. The first
one, at $F \approx F_{sk}=0.08F_{f0}$, corresponds to the situation where
the {\em superkinks} start to slide \cite{nl-mob}, whereas the second step,
occurring at $F \approx 0.18F_{f0}$ for $\eta = 0.12 \omega_x$, corresponds
to the transition of the {\em trivial} kinks to the running state, as can be
seen from the phase diagram in Fig.~\ref{rec23}. At higher forces the
increase of the mobility is smoother, and the system again reaches
intermediate {\em traffic jam} regime with $B=B_m$, where the mobility
depends on the coverage as $B_m(\theta) \propto B_f\, \,(1 -\theta)/\theta$,
$B_f=1/m\eta$ being the maximum mobility \cite{nl-mob}.

The dynamical phase diagram $(F, \eta)$ for the complex coverage 
$\theta=21/31$ is shown in the Fig.~\ref{rec23}, where we plot $F_{tk}$, 
$F_f$, $F_b$ (as in Fig.~\ref{rec12}) and, additionally, $F_{sk}$. However, 
$F_{pair}$ is not plotted because the system goes to the ``traffic jam''
regime quite smoothly, and therefore it is difficult to resolve $F_{pair}$.

One can see first, that the critical force for superkinks $F_{sk}$ is
independent on friction, while the force $F_{tk}$ increases with the
friction $\eta$ and reaches the constant value $F_{tk}\approx 0.24F_{f0}$ at
high friction, which almost exactly matches the value $F_{tk}\approx
0.23F_{f0}$ for the case of $\theta = 21/40$ coverage.

The noticeable difference with the case of $\theta = 21/40$ coverage is in
the dependence of the forward critical force $F_f(\eta)$. {\em First}, as
the value of $F_{sk}$ is very low, the low-friction region with 
$F_f = $const is very narrow ($\eta < 0.01 \omega_x$), and was not resolved in the
simulation. {\em Second}, the initial linear increase of $F_f$ with $\eta$
has a larger slope than for the $\theta = 21/40$ case; at high frictions the
critical force $F_f$ also tends to a constant value $F_f^{\ast} \approx
F_{f0}$. {\em Third}, there are some irregularities in the $F_f (\eta)$
dependence at intermediate frictions. Namely, from the plots $B(F)$ at
higher frictions (similar to the case plotted in Fig.~\ref{rbf}(b)) we found
that in the ``traffic jam'' regime the mobility does not have a constant
value $B_m$ as it is for lower frictions (the last plateau in 
Fig.~\ref{rbf}(b)), but at certain intervals of frictions $B(F)$ exhibits an additional
plateau with a larger value of the mobility, $B=B_m^{^{\prime}}$ ($B_m <
B_m^{^{\prime}}<B_f$). This should result in a lower value of the critical
force $F_f$ due to the larger kinetic energy of the system in this
intermediate state (see discussion in the next section).

Finally, we observed that the ``backward'' critical force $F_b$ is almost
independent on coverage, and its dependence on friction is again well
described by the corresponding expression for noninteracting atoms.

\section{Discussion}

\label{pheno}

The most interesting feature of both phase diagrams in Figs.~\ref{rec12} 
and~\ref{rec23} is that the critical forces, separating various intermediate
stages during the forward transition, are {\em friction dependent}, except
the first critical force, which corresponds to the completely locked 
{\em preceding} stage. From this, we may conclude that the kinetic energy of the
system in the {\em preceding} stage defines the transition to the 
{\em following} stage. Below we present a discussion which allows to rationalize
the contribution of the kinetic energy of the running kinks to the dynamical
phase transition towards the running state of atoms.

First let us consider the lowest interval of frictions, $\eta < 0.1 \omega_x$
in the phase diagrams of Figs.~\ref{rec12} and~\ref{rec23}. The explanation
of the dependence of $F_f$ on friction $\eta$ in this region can be done
solely with the help of kinetic arguments in the following way. Assume that
there is a ``critical'' velocity of kink $v_c$, above which the {\em running}
kink cannot exist as a stable quasiparticle (as was found by Peyrard and
Kruskal \cite{Michel-D}, such a critical velocity does exist for the running
kinks in the frictionless case). Then, if at given values of the force $F$
and the friction $\eta $ the kink drift velocity $\langle v_k \rangle =
F/m_k \eta_k$ is higher than $v_c$, the kink should destroy itself as soon
as it starts to move, this immediately will cause an avalanche driving the
whole system to the totally running state of atoms.

If one makes simplifying assumption that $\eta_k \sim \eta$, the drift
velocity of kinks is $\langle v_k \rangle \propto F/ \eta$ and, therefore,
the region in the phase diagram $(F,\eta )$, where the running kinks are
stable, is bounded by the straight line $F \propto v_c \eta$. This simple
linear dependence describes quite well the dependence $F_f(\eta)$ for $\eta
< 0.1 \omega_x$ (see dotted lines in Figs.~\ref{rec12},~\ref{rec23}). Thus,
for instance, in the $\theta =21/40$ case (Fig.~\ref{rec12}) for $\eta <
0.05 \omega_x$, when the applied force exceeds the threshold $F_{tk}$, the
stationary drift velocity for the kink is higher than $v_c$, and the system
goes directly to the running state of atoms. This mechanism provides, that 
$F_f=F_{tk}=$ const in this region. However, at higher frictions, $\eta >
0.05 \omega_x$, the kink can move as a stable quasiparticle, and the
transition to the running state takes place at a higher force $F_f \propto
\eta$ (Fig.~\ref{rec12}). Such a linear dependence holds also (and even
better) for the coverage $\theta=21/31$ at not too high friction 
(see Fig.~\ref{rec23}). Unfortunately, we failed to find the value of the critical
velocity explicitly, because we do not know the values for the kink mass $m_k
$ and the friction coefficient $\eta_k$. However, we observed that this
critical velocity $v_c$ is proportional to the sound velocity $c$ (defined
as $c^2 \approx a_{A}^2 V^{\prime\prime}_{int}(a_{A})$, where 
$a_{A}=a_{sx}/\theta_0$ is the average interatomic distance along the chain).
Indeed, from the slopes of dotted lines in Figs.~\ref{rec12},~\ref{rec23}
the ratio of the critical velocities for both studied coverages is 
$v_c^{(21/40)} / v_c^{(21/31)} \approx 0.72 \pm 0.07$, and the corresponding
ratio of the sound velocities for the background coverages is 
$c^{(1/2)}/c^{(2/3)}$ = 0.74.

At higher frictions, $\eta >0.1\omega _x$, $F_f$ starts to deviate
significantly from the simple linear law, and finally tends to a constant
value $F_f^{*}$. This behavior may be qualitatively explained if we take
into account that the {\em running} kinks stimulate the transition of the
system to the totally running state. Indeed, owing to their nonzero kinetic
energy $T_{kink}$, the running kinks effectively reduce the average energy
barrier for the transition of all the atoms to the running state; therefore
the effective barrier $\varepsilon _{eff}$ should be lower than $\varepsilon
_{sx}$. Let us approximate the effective barrier as $\varepsilon
_{eff}=\varepsilon _{sx}-T_{kink}$, where $T_{kink}\propto \langle
v_k\rangle ^2\propto F^2/\eta ^2$. Then, the critical force for the
transition to the running state, $F_f=C\pi \varepsilon _{eff}/a_{sx}$, is
determined by the following equation,
\begin{equation}
\label{forward}\frac \alpha {\eta ^2}F_f^2+F_f=F_f^{*},
\end{equation}
where $\alpha >0$ is a phenomenological coefficient, and $F_f^{*}\approx
F_{f0}\equiv C\pi \varepsilon _{sx}/a_{sx}$. Eq.~(\ref{forward}) provides a
qualitative agreement with the simulation data: at small $\eta $ the forward
force is proportional to $\eta $, $F_f(\eta )=\sqrt{\frac{F_f^{*}}\alpha }\eta $, 
while at $\eta \rightarrow \infty $ we have $F_f\rightarrow F_f^{*}$. 
Unfortunately, the quantitative agreement is not satisfactory. Even if 
Eq.(\ref{forward}) can provide a rather good fitting of the $F_f(\eta )$
dependence in Fig.~\ref{rec12}, the nonmonotonic peculiarities of $F_f(\eta )
$ in Fig.~\ref{rec23} remains unexplained. We recall, however, that the main
simplification made above, is that the friction coefficient for the running
kinks $\eta _k$ is equal to $\eta $. But the running kink experiences also
the {\em intrinsic} friction $\eta _{int}$, so that the correct expression
is $\eta _k=\eta +\eta _{int}$, where $\eta _{int}$ depends very nonlinearly
and {\em resonantly} on the kink velocity $\langle v_k\rangle $ \cite
{Michel-D,OMreport,Watanabe}. The intrinsic part of friction arises because
the propagating kink excites small-amplitude waves (phonons) in the discrete
atomic chain. At certain drift velocities the translational motion of kink
(i.e. the frequency with which kinks ``hits'' the particles) can come in a
resonance with the allowed phonon frequencies, which results in bigger
losses of energy by kink and increases effective damping. This is very
likely why one can observe nonmonotonic variation of $F_f(\eta )$ in 
Fig.~\ref{rec23}.

The proposed qualitative picture may be applied not only to $F_f$, but also
to other intermediate critical forces, such as $F_{pair}$ or $F_{tk}$ in the
case of the complex background $\theta =21/31$ coverage. For example, the
critical force for the trivial kinks $F_{tk}$ increases with the friction in
the $\theta =21/31$ case, when the preceding stage is characterized by the
running superkinks (Fig.~\ref{rec23}); $F_{tk}$ tends to the constant value
found for trivial background coverage $\theta =21/40$ only at high
frictions, when the kinetic energy contribution of superkinks to the
initialization of the trivial kinks motion becomes unessential.

Finally, let us comment the fact, that the asymptotic ($\eta \rightarrow
\infty $) forward critical force $F_f^{*}$, for both studied coverages, has
been found equal to the critical force for a single atom $F_{f0}=C\pi
\varepsilon _{sx}/a_{sx}$. This is easy to understand if one takes into
account that the studied atomic chain corresponds to a low-coupling limit
\cite{bklast}, because the dimensionless elastic constant 
$g_{eff}={\frac{a_{sx}^2}{2\pi ^2\varepsilon _{sx}}}\ V_{int}^{\prime \prime }(a_A)$ 
is well
below unity for both studied coverages, namely $g_{eff}\in [0.06,0.2]$ for 
$\theta \in [1/2,2/3]$. Therefore, the interaction between atoms should not
lead to a significant change of the barrier for the transfer of an atom to
the running state {\em unless the running kinks contribute to this
transition.}

\section{Conclusions}

\label{conclusions}

Thus, we have studied the dynamical phase transition from the locked to the
running state of interacting atoms in a periodic external potential under
the action of a dc external force in the underdamped limit of a generalized
Frenkel-Kontorova model. This transition proceeds by a complex multistep
scenario, which can be treated as a hierarchy of depinnings of quasiparticle
excitations of the FK model (kinks). The interesting feature of the
transition is that the critical forces separating different intermediate
stages during the ``forward'' transition, are friction dependent, except the
first critical force, which corresponds to the transition from the
completely locked ground state. This reflects the dynamical nature of
transitions between intermediate stages, namely that the kinetic energy of
running kinks in the {\em preceding} stage significantly contributes to the
transition towards the {\em following} stage.

On the basis of simulation results, we have proposed a phenomenological
approach which qualitatively explains the observed friction dependences of
the critical forces. According to this approach, for low frictions the
critical quantity determining the criterion for the transition to the
totally running state of atoms, is the drift velocity of kinks at the
preceding stage. This approach leads to the simple phenomenological equation
(\ref{forward}) for the dependence of the forward critical force on
friction. However, for better quantitative description of the simulation
results one has to take into account the resonant interaction of the running
kinks with the phonons excited in the atomic chain \cite
{Michel-D,OMreport,Watanabe}. Work along this line is in progress.

\acknowledgments

The authors thank Michel Peyrard for very helpful comments. All members of
the Centre for Nonlinear Studies at HKBU are acknowledged for stimulating
discussions. The work of M.P. and B.H. was supported in part by the Research
Grant Council, and the Hong Kong Baptist University Faculty
Research Grants.

\begin{figure}
\caption{ The mobility $B$ versus external force $F$  for the
quasi-one-dimensional FK model: (a) for $\theta=21/40$
coverage (trivial kinks on the background of
trivial $\theta_0=1/2$ structure); (b) for $\theta=21/31$ coverage
(superkinks on the background of the complex $\theta_0=2/3$  structure).
The plus signs and the solid curve correspond to
force increasing, the cross signs and the dashed curve to force
decreasing. The arrows indicate the hysteresis.
The force $F$ is expressed in terms of  the ``forward'' force for a
single atom $F_{f0} = C \pi \varepsilon_{sx}/a_{sx}$ defined in the
Introduction.}
\label{rbf}
\end{figure}

\begin{figure}
\caption{Illustration of the atomic motion in the regime of running kinks
(a) and in the ``traffic jam'' regime (b).
Immobile atoms are denoted with gray circles,
running atoms and atoms in the kink regions are denoted with
black circles, arrows indicate the direction of atomic motion.}
\label{jam}
\end{figure}

\begin{figure}
\caption{Dynamical phase diagram in the $(F, \eta)$ plane for the
quasi-one-dimensional FK model at the $\theta=21/40$, i.e. for the
trivial kinks on the background of simple $\theta_0=1/2$ structure.
The force $F$ is scaled in the same way as in Fig.~\ref{rbf}.}
\label{rec12}
\end{figure}

\begin{figure}
\caption{Dynamical phase diagram in the $(F, \eta)$ plane for the
$\theta=21/31$ coverage.
The force $F$ is scaled in the same way as in Fig.~\ref{rbf}.}
\label{rec23}
\end{figure}

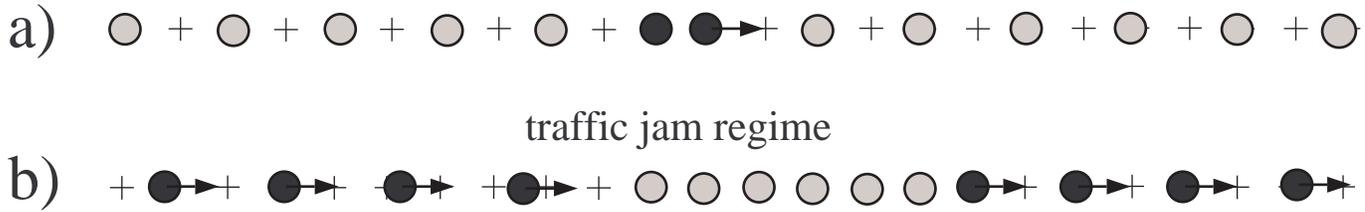
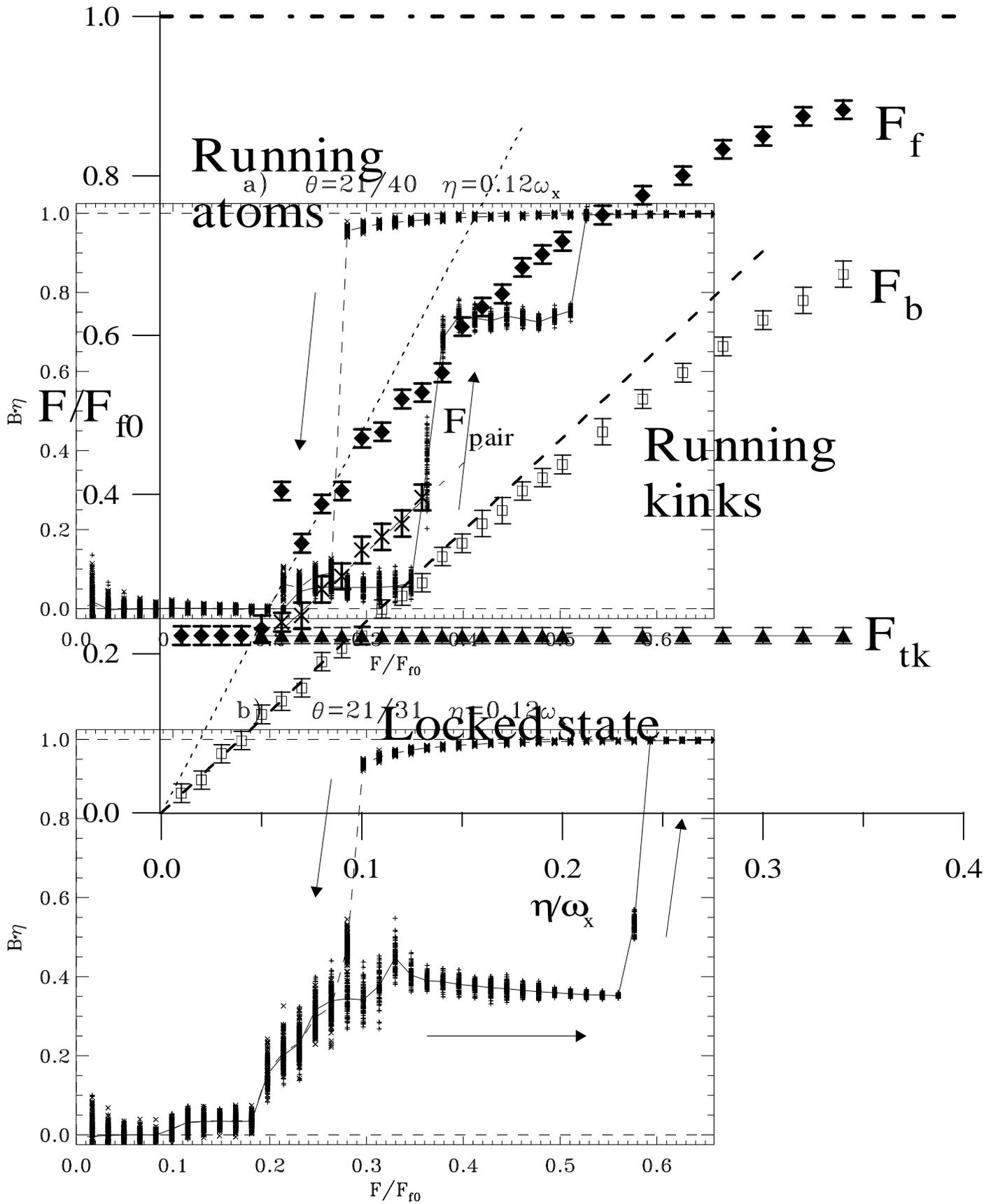
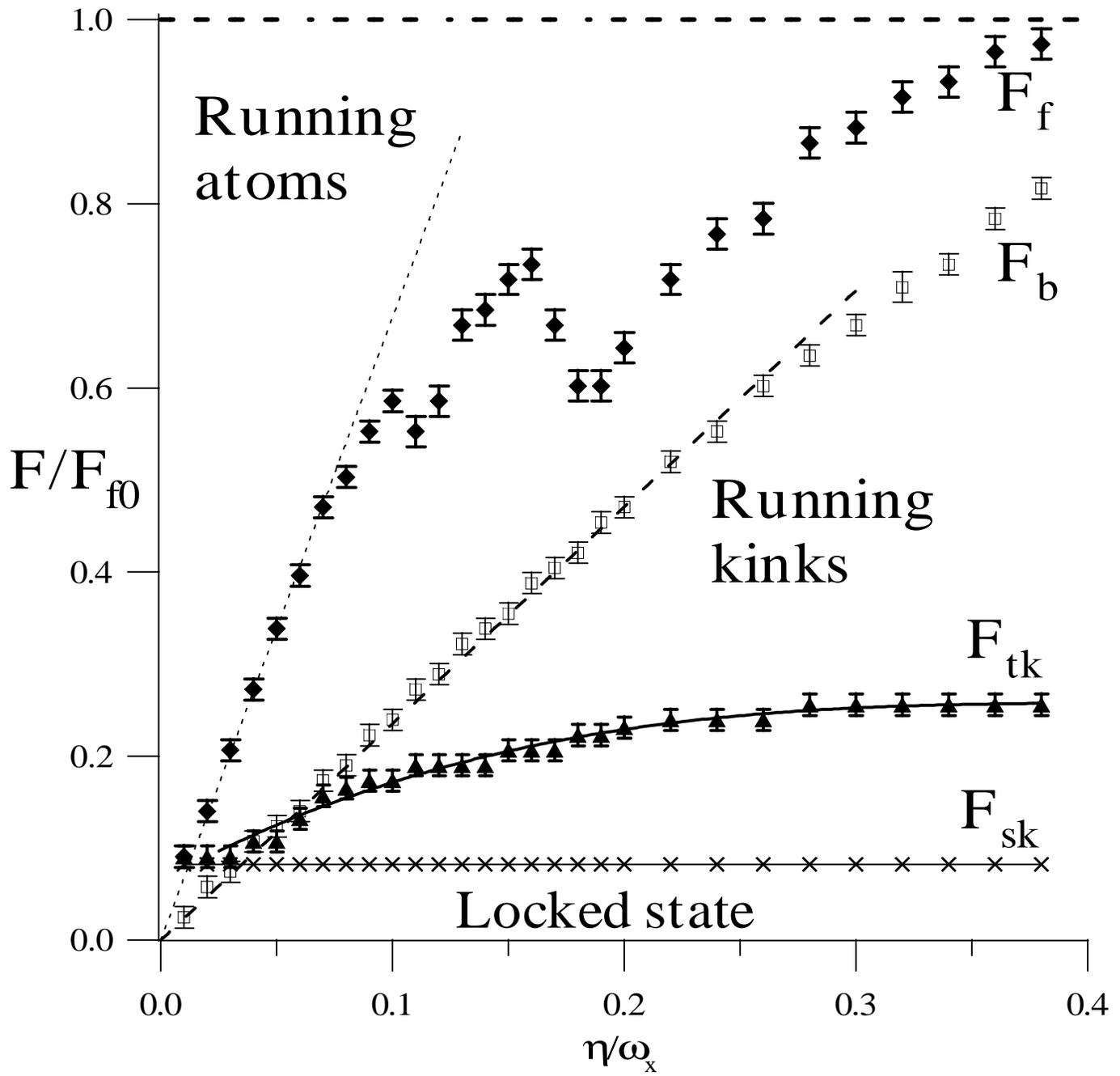
\end{document}